# Atomic structure of the unique antiferromagnetic 2/1 quasicrystal approximant

Farid Labib[1], Hiroyuki Takakura[2], Asuka Ishikawa[3] and Ryuji Tamura[1]

[1] *Department of Material Science and Technology, Tokyo University of Science, Tokyo 125-8585, Japan*
[2] *Division of Applied Physics, Faculty of Engineering, Hokkaido University, Sapporo 060-8628, Japan*
[3] *Research Institute of Science and Technology, Tokyo University of Science, Tokyo 125-8585, Japan*

The atomic structure of the newly discovered antiferromagnetic $Ga_{50}Pd_{35.5}Tb_{14.5}$ 2/1 approximant to quasicrystal with the space group of $Pa\overline{3}$ (No. 205), $a$ = 23.1449(0) Å was determined by means of a single crystal X-ray diffraction. The refined structure revealed two main building units, namely, a Tsai-type rhombic triacontahedron (RTH) cluster with three concentric inner shells and an acute rhombohedron filling the gaps in between the RTH clusters. One of the interesting findings was a very low number of chemically mixed sites in the structure, which amount to only 7.40 % of the all the atomic sites within an RTH cluster. In particular, a disorder-free environment was noticed within a nearest neighbor of an isolated $Tb^{3+}$ ion, which is presumably one of the main contributors in enhancing antiferromagnetic order in the present compound. The second significant finding was the observance of an orientationally ordered trigonal pyramid-like unit with a height of 4.2441(7) Å at the center of the RTH cluster, which has never been observed in the similar compounds before. Such unit is noticed to bring structural distortion to outer shells, in particular, to the surrounding dodecahedron cage being another possible contributor of the antiferromagnetic order establishment in the present compound. The results, therefore, are suggestive of a possible link between chemical/positional order and the antiferromagnetic order establishment.

## I. INTRODUCTION

Icosahedral quasicrystals (*i*QCs), as aperiodically-ordered intermetallic compounds, generate sharp Bragg reflections with 5-fold rotational symmetry forbidden to crystals indicating the presence of a long-range order without periodicity in their atomic configuration [1]. Among various types of classified quasicrystals, namely the Mackay- [2,3], Bergman- [4,5] and Tsai-type [6–8], the latter is the most abundant type to date. The first ever accurate structure analysis of the *i*QC has been attained in the binary $Cd_{5.7}Yb$ compound [9], thanks to its high structural quality without chemical disorder, which revealed two main building units in the structure; a multi-shell polyhedron, as demonstrated in FIG. 1a, with four concentric shells (from the outermost one): a rhombic triacontahedron (RTH), an icosidodecahedron, an icosahedron and a dodecahedron caging a central tetrahedron. Within the RTH cluster, the rare earth (*RE*) elements exclusively occupy the vertices of an icosahedron. The second building unit is an acute rhombohedron (AR) which fills the gaps between the RTH clusters and accommodates two additional *RE* elements on its long body diagonal axis.

Approximant crystals (ACs), as crystalline counterparts of *i*QCs, encompass the same building units of *i*QCs being arranged periodically with translational symmetry [10]. The term AC is commonly preceded by a rational approximation $f_{n+1}/f_n$ of $\tau$ defined as $(1+\sqrt{5})/2$ with $f_n$ representing the n-th Fibonacci number. In this scheme, the higher the rational approximation we take, the more the structure resemblances to that of an *i*QC [10]. The binary $Cd_{5.7}Yb$ *i*QC [9], $Cd_{76}Yb_{13}$ 2/1 AC [11], and $Cd_6Yb$ 1/1 AC [6] are usually believed as prototype Tsai-type compounds from which a large number of ternary or even quaternary and quinary *i*QCs and ACs have been discovered to date [12–18] by simply replacing Cd with other metallic species and Yb with various *RE* elements.

As far as their atomic structure are concerned, in the 1/1 AC, which can be described as a body centered cubic (BCC) packing of RTH clusters with the space group $Im\overline{3}$ [19], the AR unit is absent and the central tetrahedra rarely resemble to that shown in FIG. 1a. Rather, it exhibits at least one of the two classified disorder types shown in FIG. 1b, i.e., type I and type II, which refer to a 90° rotation of the tetrahedron along its two-fold axis and a positional splitting of the tetrahedron vertices, respectively [19]. There are some exceptions such as $Cd_{37}Ce_6$ and $Cd_{25}Eu_4$ 1/1 ACs, though, where the tetrahedrons are orientationally ordered with different fashions lowering the symmetry from $Im\overline{3}$ to $Pn\overline{3}$ [20] and $Fd\overline{3}$ [21], respectively. This is also worth noting that the binary $Cd_6RE$ (*RE* = Y, Sm, Gd, Tb, Dy, Ho, Er, and Tm) 1/1 ACs undergo structural transition below specific temperatures (depending on the *RE* type) where the disordered central tetrahedra become orientationally fixed [22]. Whether or not such structural transition take part in development of a long-ranged antiferromagnetic (AFM) order in the same compounds [23,24] is still an open question. In some alloy systems, referred as pseudo-Tsai type systems, on the other hand, 1/1 ACs encompass a single fully or partially occupied *RE* element at the center of the RTH cluster instead of a tetrahedron [25,26].

Despite a profound knowledge attained about the structural characteristics of the 1/1 ACs in the last two decades, little is known about the structure of their higher order counterparts such as 2/1 ACs. This might be partly related to the fact that 2/1 ACs are not as abundant as 1/1 ACs when it comes to their formation in the phase diagram. The structure of these compounds, which crystallize in the space group of $Pa\overline{3}$ [9], closely resembles to that of *i*QCs (space group: $Pm\overline{3}\overline{5}$ [9]), e.g., both comprise RTH and AR units as ther main building units. FIG. lc provides schematic view of the AR unit within a typical 2/1 AC unit cell. Based on two reports [11,27], the orientation of the innermost core unit in the 2/1 AC no longer follows the disorder modes described in FIG. 1b. Rather, three elongated arc-shaped electron density distribution, roughly resembling to a trigonal bipyramid, is noticed and explained by off-center displacement of the central tetrahedron. In one of the latest efforts in this regard [27], the

…

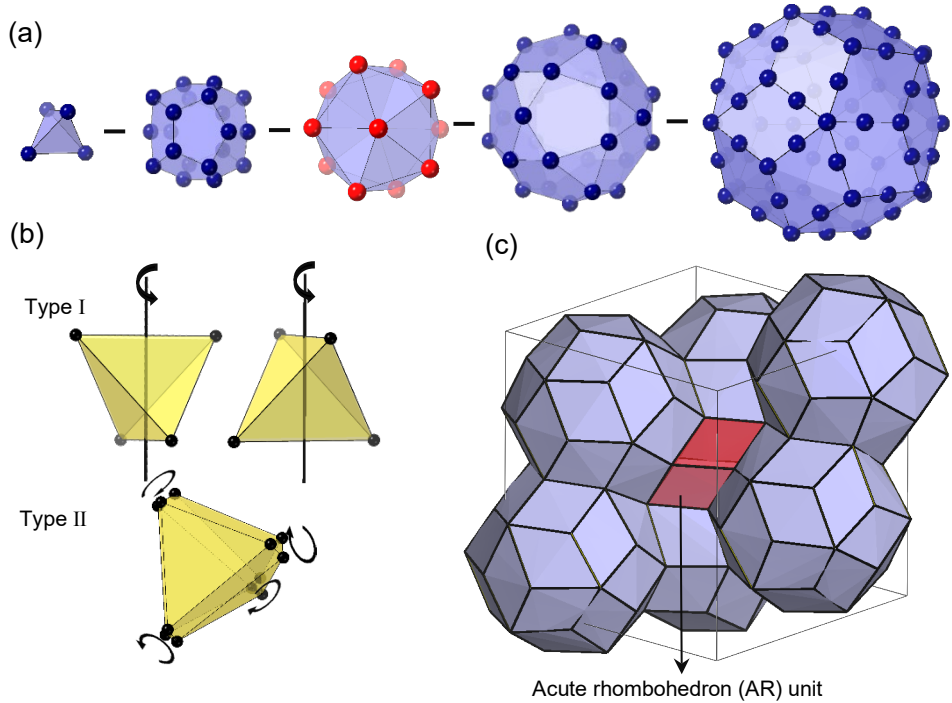

FIG. 1. (a) A typical shell structure of the Tsai-type icosahedral quasicrystal (iQC) in Cd-Tb system. From center to outermost shell: inner Cd tetrahedron (4 atoms), Cd dodecahedron (20 atoms), RE icosahedron (12 atoms), Cd icosidodecahedron (30 atoms), and Cd rhombic triacontahedron (RTH) (92 atoms). Occurrence of 12 atoms in the central tetrahedron is due to maximum splitting of the tetrahedral corner positions resulting in a cube octahedron. (b) Typical arrangement of RTH clusters within one unit cell in the 2/1 approximant (AC) including an acute rhombohedron (AR) that fills the gap between the RTH clusters.

atomic structure of the Cd-Mg-Y 2/1 AC has been examined extensively and high levels of chemical disorder between the Cd and Mg species were noticed in most of the atomic sites.

The present research, therefore, aims to carefully examine the structural parameters of the Ga-Pd-Tb 2/1 AC by means of single crystal XRD (SCXRD). This compound is of special interest due to a number of reasons. It is the only reported ternary non-Au-based Tsai-type compound that exhibits a long-ranged AFM order at low temperatures [28]. Moreover, unlike rest of the reported AFM ACs to date, this compound has high electron per atom density ($e/a$ = 1.93) under the assumption of Ga and Pd to be trivalent and zerovalent, respectively. More importantly, it is one of the two ever reported 2/1 ACs with AFM order (the other one being $Au_{65}Ga_{20}Eu_{15}$ 2/1 AC with $e/a$ = 1.54), the structure of which, in principle, contains all necessary components to construct iQC. These points further necessitate structural determination of the present compound, the outcome of which should deepen our understanding about not only the atomic structure of high-order ACs but also their structural link with corresponding iQCs, especially around the cluster center. It should also shed light on the unique AFM order observed at low temperatures and guide the quest for finding the AFM iQC in the future, which is attainable based on several theoretical studies [29–35] but yet to be discovered empirically.

## II. EXPERIMENT

Polycrystalline alloy with a nominal composition of $Ga_{50}Pd_{35.5}Tb_{14.5}$ was synthesized from high purity Ga (99.9999 wt.%), Pd (99.95 wt.%) and Tb (99.9 wt.%) elements using an arc-melting technique. The synthesized sample then heat treated at 973 K under Ar atmosphere for 120 h followed by water quenching.

Powder X-Ray diffraction (XRD) were carried out for phase identification using Rigaku SmartLab SE X-ray Diffractometer with Cu-$K_\alpha$ radiation. For structural determination, a piece of single crystal of about 0.02 × 0.03 × 0.03 mm³ in size was extracted from the sample and examined using laboratory X-ray diffraction. The single crystal X-ray diffraction (SCXRD) data was collected at room temperature using Rigaku XtaLAB Synergy-R single crystal diffractometer with a rotating anode Mo-$K_\alpha$ X-ray source. An initial structure model was obtained successfully by using SHELXT [36]. Subsequent structure refinements were conducted using SHELXL [37].

## III. RESULTS

Figure 2 represents powder XRD pattern and Le Bail fitting of the synthesized $Ga_{50}Pd_{35.5}Tb_{14.5}$ compound after an isothermal annealing at 973 K for 120 h [38]. The fitting curve was obtained by assuming the space group $Pa\bar{3}$ using Jana 2020 software suite [39]. The red, black and blue lines in the figure represent calculated ($I_{cal}$), measured ($I_{obs}$) peak intensities and the difference between them, respectively, while the black vertical bars indicate expected Bragg peak positions. As shown, there is a fair consistency between the experimental and calculated peak positions and intensities confirming high purity of the synthesized 2/1 AC. Figure 3 illustrates the reciprocal-space sections perpendicular to the [100], [110], [11$\bar{1}$], and [850] zone axes, obtained from SCXRD data after reconstruction. In total, 351711 Bragg reflections are noticed with a resolution limit equal to approximately 0.8 Å. The data reduction is performed by

…

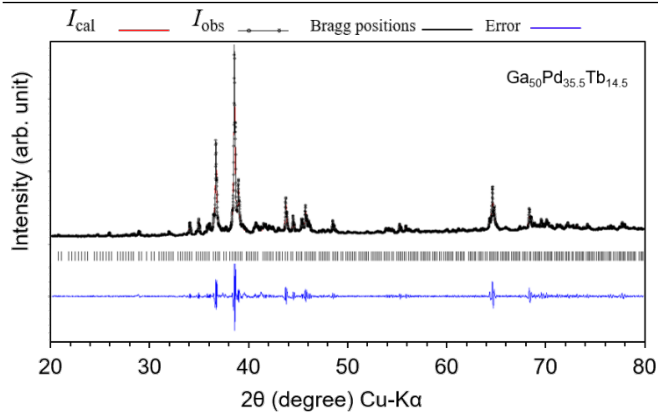

FIG. 2. Le Bail fitting of powder x-ray diffraction (XRD) patterns of the $Ga_{50}Pd_{35.5}Tb_{14.5}$ 2/1 AC annealed at 973 K.

Table I. The refinement parameters and experimental conditions

| Chemical formula | $Ga_{372.09}Pd_{264.13}Tb_{104}$ |
|---|---|
| Molar mass $M_r$ | 69409.07 |
| Temperature of data collection (K) | 298 |
| Space group | $Pa\bar{3}$ (No. 205) |
| $a$ axis (Å) | 23.1449(0) |
| Cell volume (Å³) | 12398.408 |
| $Z$ | 1 |
| $F(000)$ | 29935 |
| Calculated density (g/cm³) | 9.4518 |
| Radiation type | Mo $K\alpha$ |
| Absorption coefficient (mm⁻¹) | 43.089 |
| Crystal size (mm³) | 0.02 × 0.03 × 0.03 |
| $2\theta$ range [°] | 3.293 – 25.242 |
| No. of measured, independent and observed [I > 2σ(I)] reflections | 351711, 6319, 5875 |
| $R_{int}$ | 0.0441 |
| No. of parameters | 367 |
| No. of restrains | 0 |
| $R[F^2 > 2\sigma(F^2)]$, $wR(F^2)$, S | 0.0383, 0.0450, 1.383 |
| absorption correction | Gaussian integration |
| $T_{min}$, $T_{max}$ | 0.358, 0.479 |

assuming the space group $Pa\bar{3}$ (No. 205) leading to 5875 unique reflections and a converged reliability factor of $R\ [F^2 > 2\sigma(F^2)] = 0.0383$. The $wR(F^2)$ with $w$ defined as $1/[\sigma(F_{abs}^2)^2 + (0.0026P)^2 + 217.66P]$, equals 0.0450 for all reflections with positive intensities and the goodness of fit $S$ is 1.383. The composition of the refined model $Ga_{50.3}Pd_{35.7}Tb_{14.0}$ is in good agreement with $Ga_{50.21}Pd_{35.48}Tb_{14.31}$ obtained by ICP analysis. The refinement parameters and experimental conditions are listed in Table I. No significant violation of the systematic extinction rule for the $Pa\bar{3}$ (i.e., $0kl$: $k = 2n$, $h00$: $h = 2n$) is noticed. Table II lists atomic coordinates, Wyckoff positions, site occupations, and equivalent isotropic displacement parameters ($U_{eq.}$) after refinement.

Figure 4a presents the final refined structure model within a unit cell with a lattice parameter of 23.1449(0) Å viewed along the [100] direction where Ga, Pd, Tb atoms and unoccupied sites are represented by blue, green, red, and white spheres, respectively. Figure 4b illustrates the concentric atomic shells that constitute one RTH cluster. As a common feature in Tsai-type compounds, the RTH shell (the unit number (4)) is composed of three inner shells (assigned as units' number (1–3)) surrounding a central unit labeled as a unit (0) in FIG. 4b. The figure is quite revealing that the majority of the atomic sites (apart from a small number of disordered positions mostly around the cluster center) are fully occupied by distinct atoms. Indeed, only 7.40 % of atomic sites within one RTH

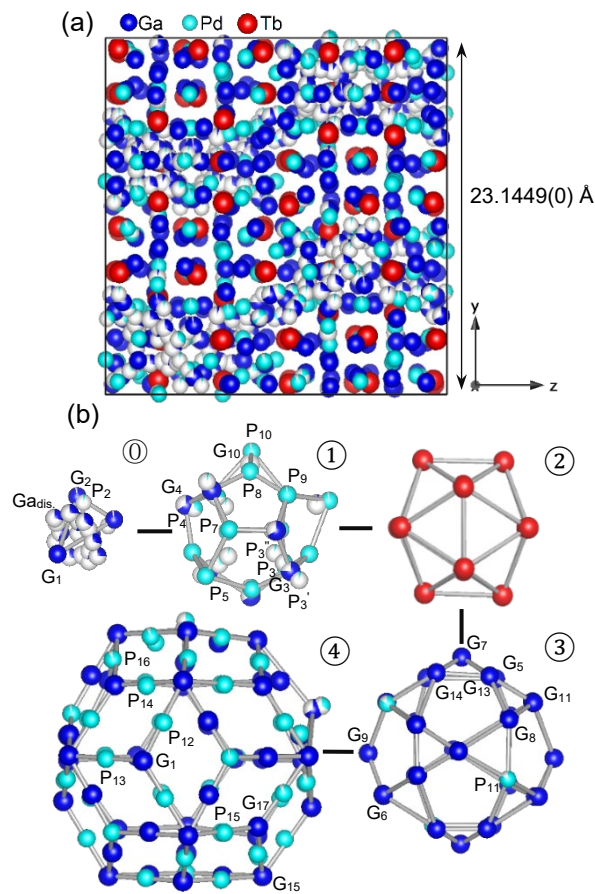

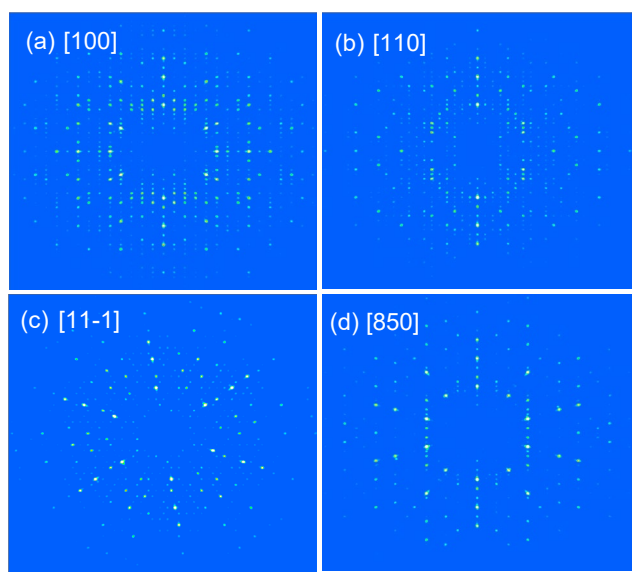

FIG. 3. Constructed reciprocal-space sections perpendicular to (a) [110], (b) [110], (c) [11-1], and (d) [850] directions of the $Ga_{50}Pd_{35.5}Tb_{14.5}$ 2/1 AC.

FIG. 4. (a) Atomic decoration within a unit cell of the $Ga_{50}Pd_{35.5}Tb_{14.5}$ 2/1 AC with a lattice parameter of 23.1449(0) Å. (b) The successive sequence of atomic shells; the inner most unit labelled as (0) wherein orientationally fixed trigonal pyramid-like unit encircled by a few translationally disordered Ga atoms, denoted as "$Ga_{dis}$" atoms, is noticed. The outer shells surrounding the central unit are (from the closest shell to the central unit): significantly deformed Ga/Pd dodecahedron, a Tb icosahedron, a Ga/Pd icosidodecahedron and a Ga/Pd rhombic triacontahedron (RTH).

…

Table II. Atomic coordinates, Wyckoff position, site occupation, and equivalent isotropic displacement parameter ($U_{eq}$) after refinement

| Unit/shell | Atom | Site | Wyck. | X | Y | Z | Occ. | $U_{eq}$ (Å)$^2$ |
|---|---|---|---|---|---|---|---|---|
| Central unit | Ga | $G_1$ | 8c | 0.0955 | 0.0955 | 0.0955 | 1 | 0.01 |
|  | Ga | $G_2$ | 24d | 0.15256 | 0.19301 | 0.25855 | 0.953 | 0.013 |
|  | Pd | $P_2$ | 24d | 0.1704 | 0.1752 | 0.2413 | 0.027 | 0.021 |
|  | Ga | Ga$_{dis.1}$ | 24d | 0.10364 | 0.18724 | 0.15696 | 0.526 | 0.008 |
|  | Ga | Ga$_{dis.2}$ | 24d | 0.0756 | 0.18645 | 0.1727 | 0.315 | 0.017 |
|  | Ga | Ga$_{dis.3}$ | 24d | 0.0871 | 0.13046 | 0.19707 | 0.274 | 0.015 |
|  | Ga | Ga$_{dis.4}$ | 24d | 0.09006 | 0.2109 | 0.1307 | 0.241 | 0.03 |
|  | Ga | Ga$_{dis.5}$ | 24d | 0.0928 | 0.1657 | 0.1844 | 0.079 | 0.053 |
| Dodecahedron | Pd | P1 | 8c | 0.1533 | 0.1533 | 0.1533 | 0.017 | 0.02 |
|  | Pd | $P_5$ | 8c | 0.03558 | 0.03558 | 0.03558 | 1 | 0.011 |
|  | Pd | $P_6$ | 24d | 0.15615 | 0.23196 | 0.35998 | 0.965 | 0.018 |
|  | Pd | $P_7$ | 24d | 0.00049 | 0.15689 | 0.09776 | 1 | 0.011 |
|  | Pd | $P_8$ | 24d | 0.09308 | 0.30819 | 0.15185 | 1 | 0.013 |
|  | Pd | $P_9$ | 24d | 0.05458 | 0.24717 | 0.25103 | 0.975 | 0.014 |
|  | Pd | $P_{10}$ | 24d | 0.15615 | 0.23196 | 0.35998 | 0.965 | 0.018 |
|  | Ga | $G_{10}$ | 24d | 0.152 | 0.2216 | 0.3274 | 0.034 | 0.008 |
|  | Ga | $G_4$ | 24d | 0.15362 | 0.27944 | 0.48278 | 0.685 | 0.008 |
|  | Pd | $P_4$ | 24d | 0.15401 | 0.2879 | 0.5048 | 0.315 | 0.026 |
|  | Ga | $G_3$ | 24d | 0.048 | 0.259 | 0.0515 | 0.73 | 0.034 |
|  | Pd | $P_3$ | 24d | 0.0703 | 0.2409 | 0.0655 | 0.244 | 0.012 |
|  | Pd | $P_3'$ | 24d | 0.0247 | 0.281 | 0.0389 | 0.026 | 0.017 |
|  | Pd | $P_3''$ | 24d | 0.0946 | 0.2167 | 0.0869 | 0.028 | 0.012 |
| Icosahedron | Tb | $Tb_1$ | 24d | 0.15257 | 0.33563 | 0.26729 | 1 | 0.007 |
|  | Tb | $Tb_2$ | 24d | 0.04149 | 0.15408 | 0.34227 | 1 | 0.008 |
|  | Tb | $Tb_3$ | 24d | 0.03748 | 0.3449 | 0.47105 | 1 | 0.008 |
|  | Tb | $Tb_4$ | 8c | 0.46049 | 0.46049 | 0.46049 | 1 | 0.008 |
|  | Tb | $Tb_5$ | 24d | 0.27219 | 0.34569 | 0.46669 | 1 | 0.01 |
| Icosidodecahedron | Ga | $G_5$ | 24d | 0.24957 | 0.28809 | 0.3514 | 1 | 0.01 |
|  | Ga | $G_6$ | 24d | 0.08683 | 0.46557 | 0.43929 | 1 | 0.01 |
|  | Ga | $G_7$ | 24d | 0.1336 | 0.4107 | 0.15624 | 1 | 0.012 |
|  | Ga | $G_8$ | 24d | 0.21345 | 0.44986 | 0.23976 | 1 | 0.011 |
|  | Ga | $G_9$ | 24d | 0.15293 | 0.35258 | 0.39839 | 1 | 0.011 |
|  | Ga | $G_{11}$ | 24d | 0.06003 | 0.28574 | 0.35412 | 1 | 0.012 |
|  | Ga | $G_{12}$ | 24d | 0.22556 | 0.46872 | 0.44028 | 1 | 0.017 |
|  | Ga | $G_{13}$ | 24d | 0.02868 | 0.35884 | 0.23294 | 1 | 0.018 |
|  | Ga | $G_{14}$ | 24d | 0.02854 | 0.36777 | 0.08628 | 1 | 0.02 |
|  | Pd | $P_{11}$ | 24d | 0.0764 | 0.215 | 0.4529 | 0.84 | 0.004 |
| RTH | Ga | $G_{15}$ | 8c | 0.35225 | 0.35225 | 0.35225 | 1 | 0.009 |
|  | Ga | $G_{16}$ | 24d | 0.14724 | 0.45567 | 0.33904 | 1 | 0.01 |
|  | Ga | $G_{17}$ | 24d | 0.0287 | 0.47858 | 0.15906 | 1 | 0.019 |
|  | Pd | $P_{12}$ | 24d | 0.24551 | 0.40255 | 0.3428 | 1 | 0.01 |
|  | Pd | $P_{13}$ | 24d | 0.05127 | 0.39913 | 0.34751 | 1 | 0.009 |
|  | Pd | $P_{14}$ | 24d | 0.33958 | 0.44681 | 0.40967 | 1 | 0.013 |
|  | Pd | $P_{15}$ | 24d | 0.09038 | 0.46338 | 0.24624 | 1 | 0.014 |
|  | Pd | $P_{16}$ | 24d | 0.06428 | 0.09295 | 0.46608 | 0.913 | 0.015 |

* For abbreviate purpose, largely displaced sites on the dodecahedron vertices are denoted as "L", while the split sites on the same sell are denoted as "S".

cluster are fractionally occupied which is exceptional among all the ternary higher order ACs reported to date. Take Cd-Mg-Y 2/1 AC as an example [27], wherein the percentage of mixed sites within the cluster amount to ~73.45 %. It seems that very low chemical disorder is a characteristic feature of the present structure. In the following, we discuss structural features of each shell within the RTH cluster starting from the innermost unit.

The core unit, assigned as a unit number (0) in FIG. 4b, appears in a curious fashion and bears no resemblance to commonly observed central units in Tsai-type or even pseudo-Tsai-type systems where, tetrahedron (either ordered or disordered) and single fully/partially occupied *RE* atom are found, respectively [26]. It rather exhibits an orientationally fixed trigonal pyramid-like unit encircled by a few positionally disordered atoms around its mid-height, which are referred as "Ga$_{dis.}$" throughout this paper. Figure 5 provides a clearer illustration of the electron density distribution at the center of the cluster by projecting 2D electron density maps on three cut planes perpendicular to [111] axis labeled as 1, 2 and 3, which bisect an iso-volume electron density cube shown in the down-left side of FIG. 6b. Planes number 1, 2 and 3 pass through a tip, a mid-height, and a base triangle of the pyramid, respectively. Clearly, there are four strong localized electron densities at the vertices of the trigonal pyramid-like unit and some smeared off-plane densities with lower intensities around its mid-height. The base of the pyramid is an equilateral triangle with a length of 3.0324(2) Å while its height and slant distances amount to 4.2441(7) Å and 4.5912(1) Å, respectively. Notice the long pyramid height that exceeds the typical bond lengths between Ga and/or Pd pairs which

...

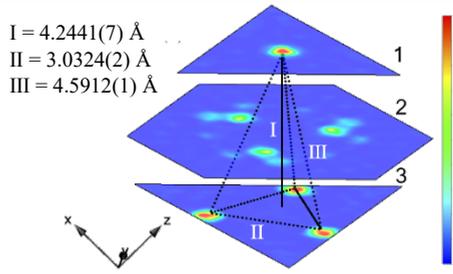

FIG. 5. (a) A 2D projections of the electron density maps on three [111] planes labeled 1, 2 and 3 passing through a tip, a mid-height and a base triangle of the trigonal pyramid, respectively.

lie in a range of 2.50-2.69 Å [40]. It can be assumed that $Ga_{dis.}$ appear to fill the extra space generated by the elongated trigonal pyramid and yet find it difficult to be stabilized due to extremely compact atomic environment. Be noted that the height of the trigonal pyramid coincides with three-fold axes of the unit cell. Whether such orientational ordering at the cluster center takes part in AFM order establishment in the present sample is unclear at the moment. In the AFM $Cd_6RE$ 1/1 ACs wherein the cubic symmetry breaks below a certain temperature due to the $Cd_4$ tetrahedra ordering, it has been proposed that the distortion of the icosahedra modifies the magnetic exchange interactions and induces the magnetic ordering possibly due to the relieve of some frustration inherent to the perfect icosahedron or octahedron arrangement of $RE$ [41]. The reason for the structural distortion in the $Cd_6RE$ 1/1 ACs is that the cubic symmetry cannot be retained after the orientational ordering of $Cd_4$ tetrahedra and thus the structure transforms to a monoclinic [42].

On the basis of above discussion, it is crucial to explore whether or not the ordering of the cluster center in the present compound is compatible with the symmetry constraints posed by the space group of $Pa\bar{3}$. Figures 6a, b and c provide spatial distribution of the isolated trigonal pyramids within a unit cell viewed along main zone axes. The red vectors represent [111] and equivalent directions. The "$a$-glide planes" and the symmetry centers (or inversion points) at (0 0 0), (½ 0 0), (½ ½ 0), (0 ½ 0) (½ ½ ½) associated with the space group $Pa\bar{3}$ are also shown in FIG. 6. Clearly, unlike 1/1 ACs, the center of cubic symmetry in the 2/1 AC does not coincide with the center of RTH cluster and thus the distribution of trigonal pyramids, the tips of which always point toward [111] directions, perfectly follows the symmetry constraints posed by the space group $Pa\bar{3}$.

As mentioned earlier, the height and the slant lengths of the pyramid are 4.2441(7) Å and 4.5912(1) Å, respectively. Allocating a unit with such dimensions especially at the center of a highly compact RTH cluster is expected to have some serious consequences to the whole structure, in particular, the surrounding dodecahedron cage. Figure 7 shows the electron density iso-surface generated from $F_{obs}$ at 18 $e$/Å$^{-3}$ level at the center of a cluster including the central unit (represented by red color) and the surrounding dodecahedron. Overall, two consequences can be distinguished on the dodecahedron shell. First is the significant displacement of the four vertices labelled as "L" sites (atoms $P_5$ and $P_6$ in Table II), which seems a reasonable measure taken to avoid too short interatomic distances between the atoms on the trigonal pyramid and dodecahedron vertices or even their collision. For example, the displacement of $P_5$ atom from its ideal position is approximately 1.67 Å. Such a displacement allows minimum distances between the trigonal pyramid and the dodecahedron vertices to be kept in a reasonable range of 2.4021(9) – 2.6186(9) Å. The resultant dodecahedron, therefore, is not a regular but a substantially deformed one. The second noticeable consequence is the site splitting of six atoms on the dodecahedron vertices marked by

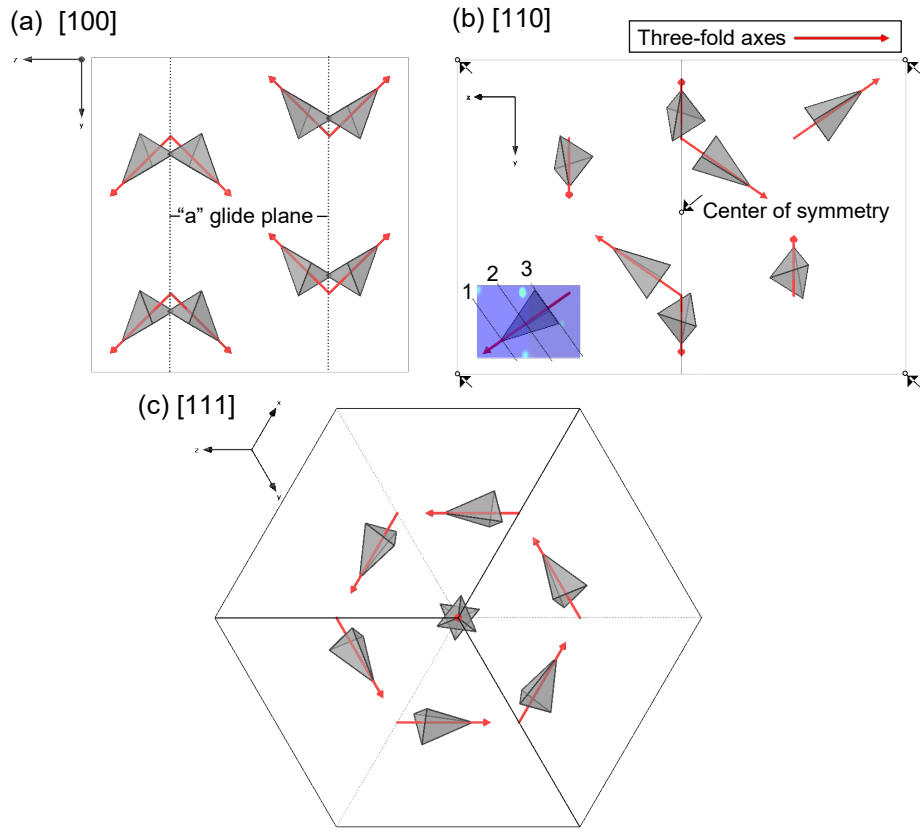

FIG. 6. The distribution of eight trigonal pyramids within one unit cell viewed along (a) [100], (b) [110] and (c) [111] directions. The red vectors that pass through the center of trigonal pyramids in Figures 6 a–c represent [111] and equivalent directions.

…

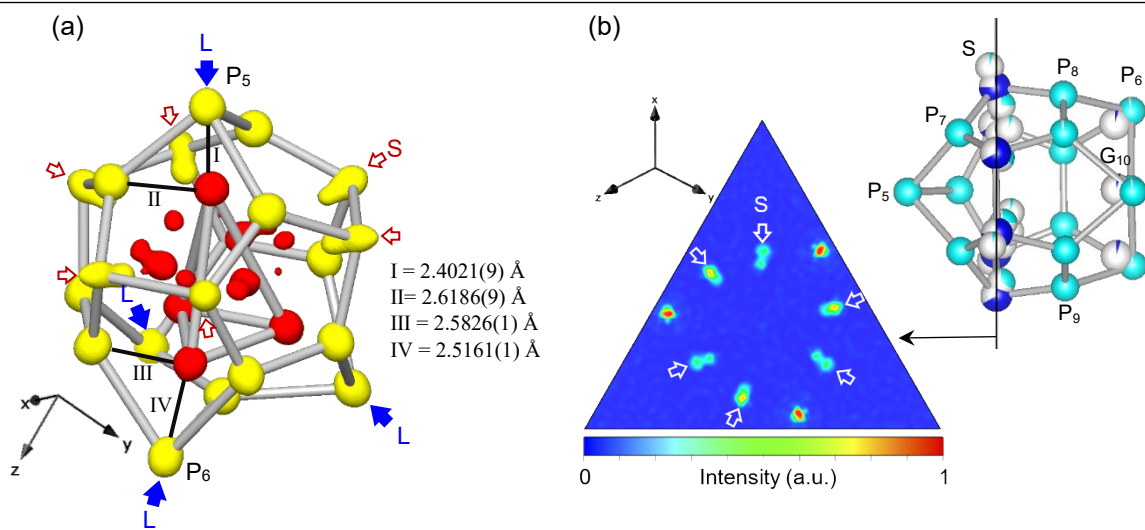

FIG. 7. (a) Electron density iso-surface at the cluster core (shown in red) and the deformed dodecahedron (shown in yellow) generated from $F_{obs}$ at 18 $e$/Å$^{-3}$ level. The blue filled arrows represent four atoms on the dodecahedron vertices that are displaced to avoid too short interatomic distances with the vertices of the triangular pyramid. (b) The 2D projections of the electron density map on [111] plane wherein the six splitting sites on the dodecahedron vertices are represented by small unfilled arrows. The three unmarked high residual electron density spots in Fig. 7b correspond to $Tb_5$ atoms in the outer shell.

unfilled red arrows in FIG. 7, hereafter referred as "S" sites. These sites, which are better visualized by projected 2D electron density map on [111] plane in FIG. 7b, are mixed positions preferentially occupied by Ga with a mixture ratio of ($Ga_{68.5-73}$:$Pd_{27.2-31.5}$). The appearance of "S" sites seems to be a natural response of the structure to the presence of $Ga_{dis.}$ atoms around the central pyramid. In other words, the disorder of the $Ga_{dis.}$ atoms extend to the nearest neighbor atoms on the dodecahedron shell in a form of positional and chemical disorder. The three unmarked high electron density spots in FIG. 7b correspond to $Tb_5$ atoms in the outer shell, which will be discussed later. Except six "S" atoms, the rest of the dodecahedron sites are dominantly occupied by Pd (see FIG. 7b).

The second shell from the center, assigned as a shell number (2) in FIG. 4b, is an icosahedron, the vertices of which are fully occupied by four different symmetrically equivalent Tb atoms, as shown in FIG. 8a. There are additional Tb atoms that appear on a long body diagonal of an acute rhombohedron (AR) unit (also shown in FIG. 8a) that fills the gap between the RTH clusters at the center of unit cell (see FIG. 1b for its schematic illustration). The local structure around an isolated $Tb^{3+}$ atoms of $Tb_1$ and $Tb_4$ types are provided in FIG. 8b and c, respectively. In the case of $Tb_1$, the surrounding non-

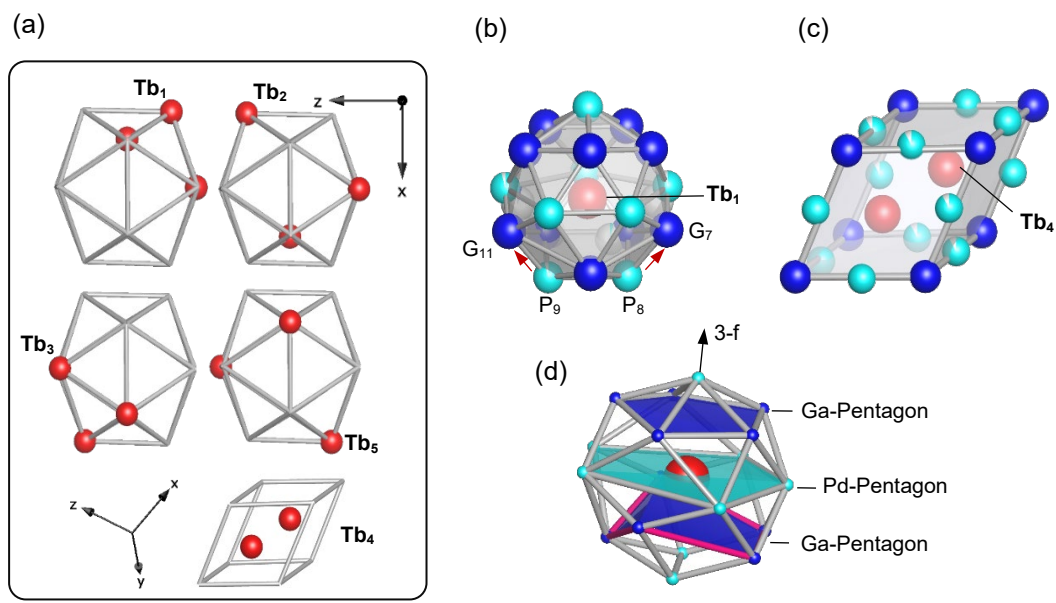

FIG. 8. (a) The Tb atoms with five different Wyckoff positions. $Tb_{1-3}$ and $Tb_5$ belong to the vertices of the icosahedral shell while two $Tb_4$ atoms appear on the long body diagonal axis of the acute rhombohedron (AR) unit that fills in the gap between the RTH clusters. (b,c) The local structure at the nearest neighbor of an isolated $Tb^{3+}$ ion located at the (b) icosahedron vertices and (c) inside the AR units. The Ga, Pd and Tb atoms are represented by dark blue, light blue and red colours, respectively. In (b), the surrounding non-Tb atoms form a $(Ga:Pd)_{18}Tb$ polyhedron with eighteen vertices. (d) The same $(Ga:Pd)_{18}Tb$ polyhedron wherein the distortion of one of the three pentagon planes due to the existence of two Pd atoms at the bottom is signified.

…

Tb atoms form a (Ga:Pd)$_{18}$Tb polyhedron with eighteen vertices, ten of which are preferentially occupied by Ga and the rest by Pd. This polyhedron is slightly different from a Cd$_{16}$RE mono-capped, double, pentagonal antiprism-like polyhedron that forms around RE elements in the Cd$_6$RE 1/1 ACs [19,20] and the difference lies in the presence of two additional vertices occupied by Pd (P$_8$ and P$_9$ in FIG. 8b) that push away the two out of five Ga atoms (G$_7$ and G$_{11}$ in FIG. 8b) on the vertices of the lower pentagonal plane leading to its out-of-plane distortion, as indicated in FIG. 8d. On the AR unit encaging the Tb$_4$ atoms shown in FIG. 8c, the eight vertices and twelve mid-edges are preferentially occupied by Ga and Pd, respectively. Interestingly, the local environments of Tb$^{3+}$ atoms on either icosahedron shell or inside the AR unit are free from chemical disorder (i.e., no mixed sites). While such a highly ordered structure around Tb$^{3+}$ atoms reflect structural perfection in the present compound, it is also suggestive of a correlation between disorder-free environment around Tb$^{3+}$ atoms and the establishment of a long-range AFM order in the present compound [28]. Indeed, the adverse effect of chemical disorder on the formation of long range AFM order in ACs has been well demonstrated by a number of researches [43,44]. A discussion on this subject, however, falls outside the scope of the present paper and is left for future studies.

The two outermost shells in FIG. 4b, denoted as shells number (3) and (4), are icosidodecahedron and RTH shells, respectively. In both shells, some degrees of distortion are noticed which seems to originate from a significant distortion of the inner dodecahedron, as discussed earlier. In addition, the atomic sites on both shells, although rarely mixed, are dominantly occupied by either Ga or Pd. In particular, among thirty vertices of an icosidodecahedron shell, twenty- seven are predominantly occupied by Ga and only three sites, denoted as P$_{11}$ in Table II, are exceptionally resided by Pd. Figure 9 displays superimposed concentric shells and projected 2D electron density map on the [13 5 8] plane where the proximity of P$_{11}$ site on the icosidodecahedron to the S sites on the dodecahedron (which are dominantly occupied by Ga) is clearly evidenced. The physical distance between these two sites by considering the averaged coordination for "S" atoms amounts to 2.497(9) Å. It seems that close proximity of these two atomic sites changes the occupational preference at the P$_{11}$ position on the icosidodecahedron shell from Ga to Pd. The unit drawn in purple in FIG. 9 will be explained in what follows.

On the outermost RTH shell, denoted as a unit number four in FIG. 4b, amongst the thirty-two vertices, twenty-six are fully composed by Ga, three vertices are mixed "S" sites and three others are fully resided by Pd. The mid-edges, on the contrary, are dominantly occupied by Pd and occasionally by Ga. In order to understand the source of a few inconsistencies in the site occupancies on the RTH shell, it is necessary to consider another building unit called an obtuse rhombohedron (OR). The OR can be described as either a shared section in between two adjacent RTH clusters or a connecting unit of adjacent dodecahedra along three-fold axes. As shown in FIG. 10a, each RTH cluster in the 2/1 AC is connected to surrounding counterparts through six two-fold and seven three-fold axes (also called $b$ and $c$-linkages, represented by thick green and yellow bonds, respectively [45]). Along the $c$-linkages, the two RTH units are interpenetrated and share sections in the shape of an obtuse rhombohedron. Another and perhaps more efficient way to describe ORs is to consider them as connecting units of adjacent dodecahedra along $c$-linkages, as shown in FIG. 10b. This way, two types of ORs denoted as OR$_I$ and OR$_{II}$, represented respectively, by blue and red polyhedron in FIG. 10 can be distinguished. The two ORs differ in their atomic arrangement and tip-to-tip length ($l$). The $l$ in OR$_I$ and OR$_{II}$ equals to 3.04 Å and 2.85 Å, respectively. Shorter $l$ in OR$_{II}$ indicates that it is rather more compressed than OR$_I$ along its height simply because both its tips are occupied by P$_5$ atoms, which correlate to the largely displaced dodecahedron "denoted as L" vertices, as discussed in FIG. 7. In OR$_I$, however, one tip corresponds to the "L" site while the other one correlates to splitting "S" sites. Since the center of OR$_{II}$ always coincides with the inversion points associated with the space group $Pa\bar{3}$ at (0 0 0), (½ 0 0), (½ ½ 0), (0 ½ 0) and (½ ½ ½), the atoms on each face of the OR$_I$ are inverted through its center. Each RTH shell is composed of seven ORs, six of which are of type I and only one is of type II. One can easily notice that all the positions on the RTH shell whose occupancies are inconsistent with the rest of the RTH sites are, indeed, those that belong to OR$_I$ and OR$_{II}$ units. Simply put, the chemical and positional disorder of some atomic sites on the dodecahedron shell affect the occupational preferences of the atomic sites on their inter-connecting OR units, which are also shared sections of neighboring RTH units. In terms of directions for future research, it will be interesting to explore possibility of eliminating Ga$_{dis.}$ atoms around an orientationally ordered central units and obtain almost defect-free structure in the ternary system. Optimizing the elemental composition and heat treatment procedure during the synthesis are possible approaches in this direction.

## IV. CONCLUSION

The present research was undertaken to investigate atomic structure of the antiferromagnetic Ga-Pd-Tb 2/1 approximant to quasicrystal by means of single crystal X-ray diffraction at room temperature. This compound is interesting because it is the only higher order approximant with antiferromagnetic order which has a relatively high electron density ($e/a$ = 1.93), where frustrated magnetic states such as spin-glass are expected. In this study, special focus was given to any uncommon structural aspects which could possibly contribute to the antiferromagnetic order establishment in this compound. The refinement of the diffraction data unveiled extremely low chemical disorder as the fractionally occupied atomic sites amounted to only 7.40 % of the atomic sites in the whole structure. Such a trivial

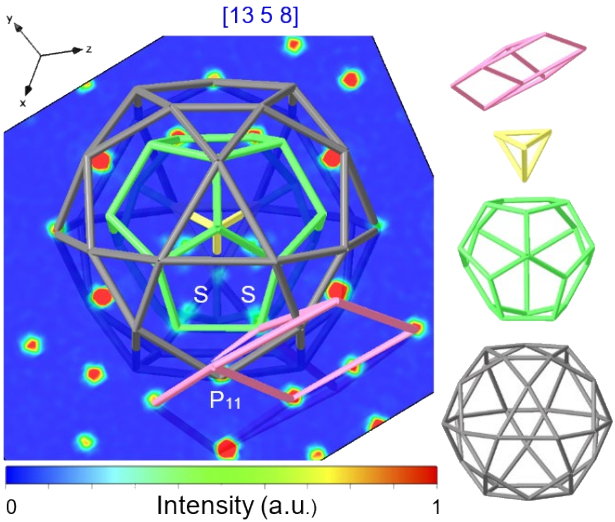

FIG. 9. Projected 2D electron density map on [13 5 8] plane passing through a P$_{11}$ site on the icosidodecahedron and the nearest dodecahedron vertices accommodating split S atoms. The figure evidences that the three vertices of the icosidodecahedron, which are exceptionally resided by Pd, locate in a vicinity of the split S atoms on the dodecahedron. The P$_{11}$ site also belongs to the inner mid-edge of an obtuse rhombohedron (OR) unit drawn in purple.

…

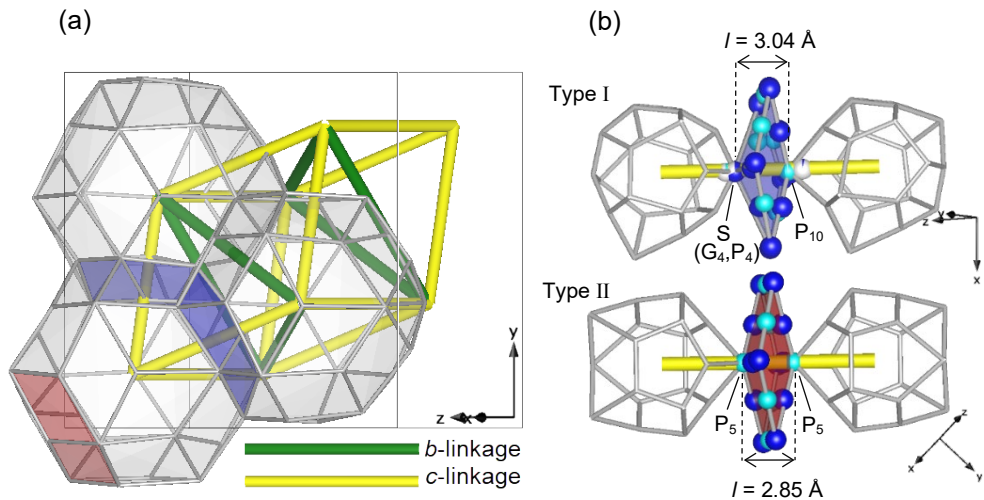

FIG. 10. (a) The arrangement of three RTH clusters within a unit cell connecting through *b* and *c*-linkages represented by thick green and yellow bonds, respectively. Along the *c*-linkages, the two RTH units are interpenetrated and share sections in the shape of obtuse rhombohedron (OR). (b) Another illustration of ORs as connecting units of adjacent dodecahedra along *c*-linkages. Two types of ORs differing in their atomic arrangement and tip-to-tip length (*l*) can be distinguished. Six out of seven ORs around each decahedron are of type I while only one is of type II.

chemical disorder is quite exceptional among other ternary higher order ACs where mixed sites within a cluster roughly amount to ~ 70 %. In particular, within a nearest neighbor of an isolated $Tb^{3+}$ ion, a disorder-free environment was noticed, which is presumably one of the main contributors in enhancing antiferromagnetic order in the present compound. Moreover, an orientationally ordered trigonal pyramid-like unit was discovered at the center of a multi-shell polyhedron, known as a rhombic triacontahedron cluster marking the first observation of such kind in higher order approximants. The orientation of the trigonal pyramids was found to perfectly follow the symmetry constraints posed by the space group $Pa\bar{3}$. Whether or not cluster center ordering contribute in establishment of a long-range magnetic order is still unclear. It is believed that the ordering brings structural distortion which might be sufficient enough to partially relive geometrical magnetic frustration and favor a long-range magnetic order. Furthermore, the surrounding dodecahedron cage of the central unit is found to be significantly distorted, in particular on its four vertices close to trigonal pyramid vertices. It was shown that the occupational preferences of atomic sites on obtuse rhombohedron units are highly affected by the chemical and positional disorders of neighboring dodecahedron sites. Taken together, the present results offer valuable insights into possible atomic scale tunning approaches to obtain antiferromagnetic order in the higher order approximants or even quasicrystals. The combination of highly ordered structure and some levels of structural distortion seem to serve as key components in developing antiferromagnetic order in the inherently frustrated Tsai-type compounds.

## ACKNOWLEDGMENT

This work was supported by Japan Society for the Promotion of Science through Grants-in-Aid for Scientific Research (Grants No. JP19H05817, No. JP19H05818, No. JP19H05819, and No. JP21H01044) and JST, CREST Grant No. JPMJCR22O3, Japan.

…

…